\documentclass[amsmath,amssymb,aps,showpacs,twocolumn,prl,floatfix]{revtex4}
\usepackage{graphicx}
\usepackage{dcolumn}
\usepackage{bm}
\usepackage{color}
\usepackage{multirow}
\usepackage{times}
\usepackage{ulem} 

\newcommand{\EF}{\ensuremath{E_{\rm F}}}

\begin{document}

\title{Giant spin Nernst effect induced by resonant scattering at surfaces of metallic films}

\author{Nguyen H. Long} 
\email{h.nguyen@fz-juelich.de}
\affiliation{Peter Gr\"unberg Institut and Institute for Advanced Simulation, 
Forschungszentrum J\"ulich and JARA, D-52425 J\"ulich, Germany}
\author{Phivos Mavropoulos}
\author{Bernd Zimmermann, Stefan Bl\"ugel} 
\affiliation{Peter Gr\"unberg Institut and Institute for Advanced Simulation, 
Forschungszentrum J\"ulich and JARA, D-52425 J\"ulich, Germany}
\author{Yuriy Mokrousov} \email{y.mokrousov@fz-juelich.de}
\affiliation{Peter Gr\"unberg Institut and Institute for Advanced Simulation, 
Forschungszentrum J\"ulich and JARA, D-52425 J\"ulich, Germany}
\date{\today}

\begin{abstract}
A new concept realizing giant spin Nernst effect in nonmagnetic metallic films is introduced. 
It is based on the idea of engineering an asymmetric energy dependence of the longitudinal and transverse electrical conductivities, as well as a pronounced energy dependence of the spin Hall angle in the vicinity of the Fermi level by the resonant impurity states at the Fermi level. 
We employ an analytical model and demonstrate the emergence of a giant spin Nernst effect in Ag(111) films using {\it ab-inito} calculations combined with the Boltzmann approach for transport properties arising from skew scattering off impurities.
\end{abstract}


\maketitle

Within the past few years, the field of spin caloric transport has attracted broad interest owing to new challenges and vistas in applications  which combine spintronic as well as thermoelectric concepts~\cite{bauer12}. 
In this field, thermal gradient is used as an ultimate agent to generate a spin current, in analogy to the generation of a charge current in conventional thermoelectrics. 
As a promiment spincaloritronics phenomenon, the relativistic spin Nernst effect (SNE) enables a way to generate a pure transverse spin current in a sample subject to an applied temperature gradient~\cite{cheng08,liu10,ma10}. 
The SNE bears an analogy to the spin Hall effect (SHE)~\cite{dyakonov71,hirsch99}, which has become one of the most efficient ways of generating spin currents in spintronics.
Owing to the fact that the SHE has been successfully observed in various types of experiments~\cite{kato04,saitoh06,kimura07}, it is expected that the spin Nernst effect would also be detectable. 
However, limitations on the magnitude of temperature gradients in metals can diminish the magnitude of the spin Nernst currents~\cite{tauber12}. 

{\it Ab-initio} studies~\cite{gradhand10,gradhand10-2,lowitzer11,gradhand12} and experiments~\cite{niimi12} suggest that the extrinsic SHE induced by the skew-scattering off impurities can be large due to a large difference in the spin-orbit coupling strength of the impurities and the host.
However, this argument is not applicable to the SNE, because the thermal transport coefficients entering the expression for the spin Nernst conductivity (SNC) are determined to a first approximation by the derivative of the conductivities around the Fermi energy ($\EF$), and not by the their values directly at it. 
As a consequence, the SNE is more sensitive to changes in the electronic structure as a function of energy, as compared to the SHE.
A requirement for the SNC to be large is that the energy dependence of the conductivities should be very asymmetric with respect to $\EF$. 

Recently, Tauber {\it et al.}~\cite{tauber12,tauber13}, using first-principles techniques combined with Boltzmann approach, computed the SNE in Cu bulk, caused by spin-dependent scattering off substitutional impurities such as Ti, Au, Bi. 
The magnitude of the SNC was predicted to be about 16\,(A/K m) at 300\,K in Cu$_{0.99}$Au$_{0.01}$ alloy. It corresponds to a spin current of about 10\,$\mu$A when using a sample with the dimensions of 100$\times$100$\times$100\,nm~\cite{kimura07} and a temperature 
gradient of 50\,K/$\mu$m~\cite{slachter10}.
For the same Cu(Au) alloy, Wimmer {\it et al.} obtained a somewhat larger value of the SNC of 30\,(A/K m) at 300\,K using an {\it ab-initio} approach based on Kubo formalism~\cite{wimmer13}. 
Although these works suggest that the corresponding magnitude of the spin current is large enough, it has not been detected so far.  
Therefore, finding systems with much larger SNE is essential for realizing the effect in experiments and utilizing it in devices. 

\begin{figure*}[t!]
\includegraphics[scale=0.5,clip=true,angle=270]{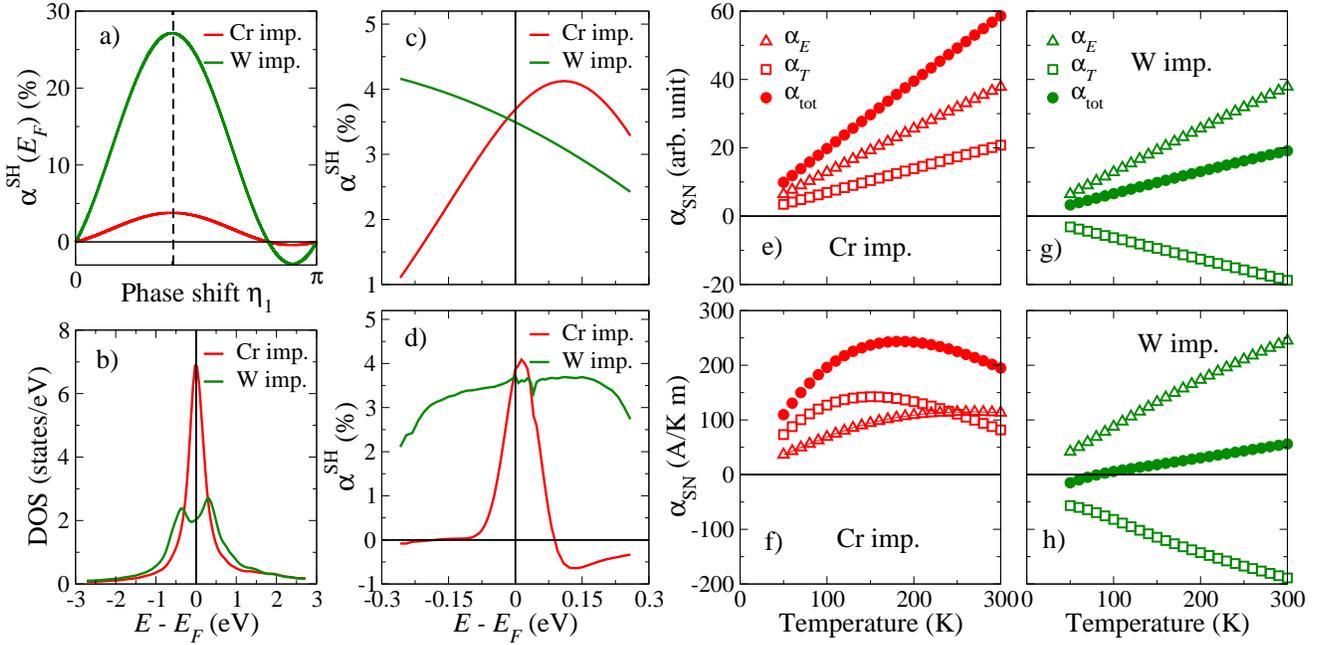}
\caption{ 
(a) The spin Hall angle at the Fermi level as a function of the phase shift of non-resonant channels according to the
model of Fert and Levi; (b) The density of states of Cr and W adatom defects deposited at the surface of a 10-layer 
Ag(111) film; (c-d) The spin Hall angle as a function of energy for W and Cr impurities according to the model of 
Fert and Levi (c) and first principles calculations (d); (e-h) Total spin Nernst conductivity $\alpha_{\rm tot}$ decomposed
into electrical $\alpha_E$ and thermal $\alpha_T$ contributions for Cr (e, f) and W (g, h) defects according to the
model of Fert and Levi (e, g) and {\it ab-initio} calculations (f, h).}
\label{figCrW}
\end{figure*}

In this work, we propose a new concept for engineering a giant spin Nernst effect in nonmagnetic metallic films by means of resonant impurity scattering. 
Previous work of Fert and Levy \cite{fert11} showed a strong influence of resonant scattering on the spin Hall effect.
Here we demonstrate that the influence of resonant scattering on the spin Nernst effect can be also remarkable. 
Namely, we analytically find that scattering off sharp resonant impurity states at surfaces of metallic films leads to a strong asymmetric energy-dependence of longitudinal charge and transverse spin Hall conductivities, as well as of the spin Hall angle. 
As a result, the spin Nernst conductivity is shown to be gigantic in a wide range of temperatures. 
Based on {\it ab-initio} calculations we demonstrate the validity of our findings by taking Ag(111) films with adatom impurities as a test system. We show that the magnitude of the SNC in this system can exceed the values reported so far by one order of magnitude. 
The formulated concepts could be used to generate large transverse spin currents with temperature gradients.

We start by summarizing the expressions we use to compute the SNC in thin metallic films~\cite{tauber12, wimmer13}, and 
then discuss the conditions for enhanced SNE. The cartesian axes are chosen such that $z$-axis is always the film normal. 
In fcc (111) films, $x$- and $y$-axes are chosen to be along [1$\bar1$0] and [$\bar1$0$\bar1$] crystallographic directions.
The temperature gradient $\nabla T$ is applied along the $x$-axis, while we look at the spin current propagating in $y$-direction with the spin-polarization along the $z$-axis.
We define the spin Nernst conductivity $\alpha_{SN}$ from the spin current density 
$j^s_{y}$ according to
\begin{equation}
j^s_y=\alpha_{SN}\nabla_xT
\label{scurrden}.
\end{equation}
The SNC comprises two contributions, $\alpha_E$ and $\alpha_T$,~i.e.,  {\it electrical} and {\it thermal} parts~\cite{tauber12}, given respectively by:
\begin{equation}\label{nernstcond}
\alpha_{\rm tot}=\alpha_E+\alpha_T =-eL^s_{0,yx}S_{xx}-(1/T)\,L^s_{1,yx},
\end{equation}
where $e$ is the electron charge and $L^s_{n,ij}$ are the $n$-th order spin transport coefficients defined below. 
The electrical part $\alpha_E$ originates in the SHE due to the internal (thermo-)electric field $E_x$ that compensates for the charge imbalance induced by $\nabla T$. 
It is given by $E_x=S_{xx}\nabla_xT$  with the Seebeck coeficient $S_{xx}=-\frac{1}{eT}\frac{L^q_{1,xx}}{L^q_{0,xx}}$. 
The thermal part of the SNC reflects the direct influence of the temperature gradient. 
The charge and spin transport coefficients, indicated by superscripts $q$ and $s$, respectively, are
\begin{equation}
L^{q(s)}_{n,xx(yx)}=\frac{1}{e}\int dE\,\sigma^{q(s)}_{xx(yx)}(E)\,\frac{\partial f_0(E,T)}{\partial E}\,(E-\EF)^n
\label{transportcoeff},
\end{equation}
where $n=0$ or 1, $f_0(E,T)$ is the Fermi distribution function, and $\sigma^{q(s)}_{xx(yx)}(E)$ are the energy dependent longitudinal charge (and transverse spin Hall) conductivities.
As apparent from Eq.~\ref{transportcoeff}, necessary condition for having large transport coefficitents and large SNC is the asymmetry of $\sigma^{q(s)}_{xx(yx)}(E)$ around the Fermi level. 

However, as follows from Eq.~\ref{nernstcond}, if $\sigma^q_{xx}$ and $\sigma^s_{yx}$ have a similar energy dependence, the electrical $\alpha_E$ and the thermal $\alpha_T$ contributions will have an opposite trend with temperature, i.e. one increases and the other decreases with increasing temperature. 
As a consequence, this will give rise to a small total SNC. 
To show this we define the energy-dependent spin Hall angle (SHA) as $\alpha^{\rm SH}(E)=\sigma^s_{yx}(E)/\sigma^q_{xx}(E)$. 
The opposite sign of the $\alpha_E$ and $\alpha_T$ can be then recast as the SHA being nearly constant in energy, $\alpha^{\rm SH}(E)=\alpha^{\rm SH}_0$, with 
\begin{equation}
\frac{L^s_{1,yx}}{L^q_{1,xx}}= \frac{L^s_{0,yx}}{L^q_{0,xx}}=\alpha^{\rm SH}_0,
\label{hallangle}
\end{equation}
as follows from Eq.~\ref{transportcoeff}. Thus, it immediately follows that 
\begin{equation}
\alpha_E=(1/T)\,\alpha^{\rm SH}_0\,L_{1,xx}=-\alpha_T.
\label{nersntcond3}
\end{equation}
We can therefore conclude that {\it strong variation of the SHA in the vicinity of the Fermi level} is a necessary condition for  $\alpha_E$ and $\alpha_T$ enhancing each other's contribution to the total SNC.

A possible scenario to realize a non-trivial energy dependence of the SHA as well as manifestly asymmetric energy dependence of the charge and spin conductivities could be by scattering off resonant impurity states positioned around the Fermi level. 
As it was pointed out by Fert and Levy \cite{fert11}, resonant scattering off $d$-impurities can enhance the spin Hall effect. 
We apply their model for SHE in terms of the energy dependence of the SHA, to the spin Nernst effect. 
Fert and Levy's expression for the SHA reads:
\begin{equation}
\alpha^{\rm SH}(E)=\frac{3}{5}\frac{\lambda_d}{\Delta}\sin{(2\eta_2(E)-\eta_1)}\sin{\eta_1},
\label{SHAE}
\end{equation}
where $\lambda_d=2/5(E_{5/2}-E_{3/2})$ is the splitting between the $d_{5/2}$ and $d_{3/2}$ impurity levels, $\Delta$ is the width of the resonance, $\eta_1$ is the phase shift of non-resonant channels and $\eta_2$ is the phase shift of resonant channels. 
Phase shift $\eta_1$ can be assumed to be energy-independent since it changes very little around the Fermi level. 
The phase shift $\eta_2$ is strongly energy-dependent and can be evaluated according to:
\begin{equation}
\eta_2(E)=\frac{2}{5}{\rm acot}\left(\frac{E_{3/2}-E}{\Delta}\right)+\frac{3}{5}{\rm acot}\left(\frac{E_{5/2}-E}{\Delta}\right),
\label{eta2}
\end{equation}
with its Fermi energy value $\eta_2(E_F)=\pi Z_d/10$, where $Z_d$ is the number of occupied impurity $d$-states.
The longitudinal charge conductivity is evaluated as $\sigma_{xx}(E)\propto 1/{\rm sin}^2(\eta_2(E))$, and the transverse spin conductivity is related to it by $\sigma_{yx}^s(E)=\sigma_{xx}(E)\cdot \alpha^{\rm SH}(E)$. 

At this point, we use first-principles calculations based on the relativistic full-potential Korringa-Kohn-Rostoker (FP-KKR) Green function method~\cite{stefanou90, heers11} for accessing the spin-dependent scattering off adatom defects at surfaces of metallic films. The particular system that we choose
to investigate is a 10-layer thick Ag(111) film. The energy-dependent longitudinal charge and transverse spin Hall conductivities due to scattering off impurities entering Eq.~\ref{transportcoeff} are calculated using the Boltzmann approach~\cite{gradhand10,herschbach12,long14}. The scattering rates as well as the conductivities are calculated 
at a nominal impurity concentration of 1\% per surface unit cell. At first, we consider two types of defects: Cr and W impurities at the adatom site positioned on the surface of Ag(111) film. 
To have a general comparison between Cr and W, Cr is considered as non-magnetic defect in this work.
Our {\it ab-initio} calculations for the density of states (DOS)
of the impurities show that while Cr DOS exhibits a very sharp resonant feature at $E_F$, the DOS of W defects
has a wider spread around the Fermi level, see Fig.~\ref{figCrW}(b).  For these two cases, we proceed both with
analytical and first principles evaluation of the SNE, as outlined above.

To estimate key parameters in the model of Fert and Levy, we take from experiment the values for $\lambda_d$ of 0.032~eV for Cr and 0.38~eV for W~\cite{vijayakumar96}. From the shape of the {\it ab-initio} DOS, Fig.~\ref{figCrW}(b), we 
estimate the resonance width $\Delta$ to be 0.49~eV for Cr and 1.3~eV for W impurities.
To estimate the phase shift of non-resonant channel, we calculate $\alpha^{\rm SH}$ at $E_F$ as a function of $\eta_1$ according to Eq.~\ref{SHAE}, with the result shown in Fig.~\ref{figCrW}(a).
It can be seen that the maximum value of $\alpha^{\rm SH}(E_F)$ is reached when $\eta_1=\eta_2=\pi Z_d/10$ with $Z_d=4$, and it can be nearly 30\% for W, with a much smaller value of only 4\% for Cr.
Using the computed from first-principles value of $\alpha^{\rm{SH}}(E_F)$ of 3.5\% both for W and Cr impurities at the surface of a 10-layer Ag(111) film, Fig.~\ref{figCrW}(d), we find that $\eta_1=0.47\pi$ for Cr defects, which is much larger than the corresponding $\eta_1=0.05\pi$ for W defects. 
This is probably a consequence of the similar extension of the non-resonant $s$-states in the W impurities compared to the Ag host.

\begin{figure}
\includegraphics[scale=0.37,clip=true,angle=270]{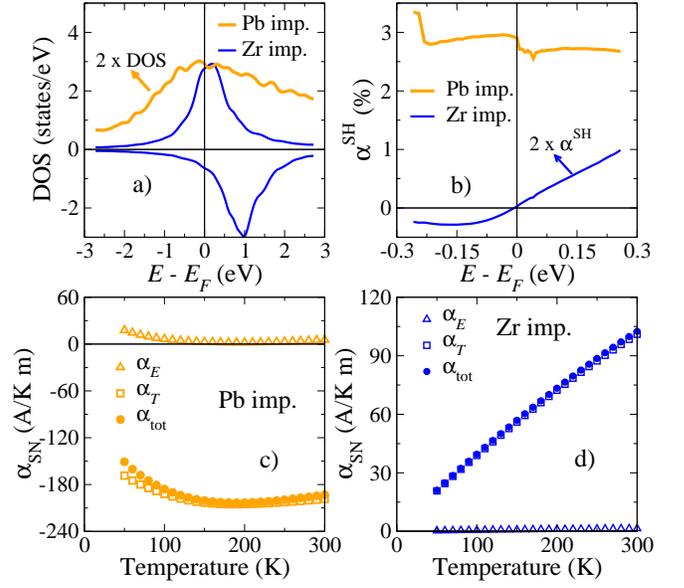}
\caption{First-principles results for the (a) density of states, (b) energy-dependence of the spin Hall angle, and the spin Nernst conductivity of Pb (c) and Zr (d) adatom defects at the surface of a 10-layer Ag(111) film.} 
\label{figPbZr}
\end{figure}

We are ready now compute the energy dependence of the spin Hall angle around the Fermi energy
using the model of Fert and Levy as given by Eq.~\ref{SHAE} \cite{footnote1}. 
It is shown in Fig.~\ref{figCrW}(c) for Cr and W impurities on top of 10-layer Ag(111) film, 
while in Fig.~\ref{figCrW}(d) the corresponding
first-principles calculations are shown. In Fig.~\ref{figCrW}(c) one can clearly see that
scattering off sharp resonant states of Cr gives rise to a very drastic variation of the SHA around the
Fermi level, even accompanied by a sign change according to {\it ab-initio} results. On the other hand, the predicted
behavior of the SHA in the case of W impurities is much smoother in energy. The difference in the qualitative
trend of the SHA between Cr and W defects, as captured within the model, is relatively well reproduced
by first principles results.  

Next, following Eq.~\ref{transportcoeff}, we integrate $\sigma^q_{xx}(E)$ and $\sigma^s_{yx}(E)$ to obtain the transport coefficients and compute the electrical $\alpha_E$ and thermal $\alpha_T$ parts of the SHE, as given by
Eq.~\ref{nernstcond}, as a function of temperature. The corresponding contributions, as well as their sum comprising
the total SNC, are shown for Cr and W defects as determined from model [Fig.~\ref{figCrW}(e, g)] and {\it ab-initio}
calculations [Fig.~\ref{figCrW}(f, h)].  
As is immediately clear from these plots, the model predictions qualitatively agree with the corresponding first principles results in terms of relative magnitude of $\alpha_E$ and $\alpha_T$, as well as their trend with temperature. 
Namely, for Cr impurities, due to the pronounced variation of the SHA around $E_F$, the electrical  and thermal parts of the SNE have the same sign, magnitude and trend with increasing temperature. The two contributions are thus added 
up to result in a very large SNC reaching as much as 250 (A/K m) at around 150 K.
In contrast, even though the two contributions in W case separately acquire values as large as those for Cr defects, they suppress each other owing to their opposite sign. We can thus attribute the small magnitude of the total SNC of about 50 (A/K m) in this case to the smooth behavior of the SHA with energy around the Fermi level. Overall, the results are 
in direct accordance to the mechanism for enhanced SNE we formulated above.

To demonstrate the emergence of large SNE due to the resonant scattering off impurities as a general phenomenon, we consider the case of the $p$-type impurity Pb and the $d$-type magnetic impurity Zr deposited on the surface of a 10-layer Ag(111) film. 
In Fig.~\ref{figPbZr}, where the first principles calculations for Pb and Zr impurities are shown, we observe that the DOS of Pb $p$-states exhibits a peak at the Fermi level, related to resonant scattering of host electrons. 
This gives rise to a drop in the SHA as the Fermi energy is crossed. 
The asymmetry in the SHA here is due to the asymmetric behavior of the spin Hall conductivity, while the energy-dependence of the longitudinal charge
conductivity is rather symmetric around $E_F$. As a result, the total SNC is dominated by the thermal contribution $\alpha_T$ of about 200 (A/K m) at 300\,K, with the electrical contribution $\alpha_E$ basically remaining negligible
in the entire interval of considered temperatures. When Pb impurities are replaced by Tl impurities with one electron less (not shown), the resonant peak is shifted far away from the Fermi level. As a consequence, $\alpha_E$ and $\alpha_T$ contributions become opposite in sign, which significantly suppresses the overall SNC in the entire temperature range. 

We also find a large spin Nernst effect for a Ag(111) film with magnetic Zr surface adatom defects, see Fig.~\ref{figPbZr}. 
In this case, the majority-spin states of Zr are positioned directly at the Fermi level, which leads to an 
asymetric behavior of the SHA as a function of energy owing to resonant scattering. 
It is remarkable that, although our calculations predict a tiny magnitude of the spin Hall angle of the order of 0.02\%, the system
exhibits a large SNC reaching as much as 115\,(A/K m) at room temperature. This serves as a perfect demostration of the fact that the SNE has little to do with the magnitude of the spin Hall effect. In turn, this
suggests that the material base for the large spin Nernst effect can be completely different from that explored currently experimentally for large SHE. 
We also have to stress that Zr impurities exhibit finite magnetic moments taken to be in the out-of-plane direction in our calculations, and 
that, correspondingly, a part of the spin Nernst conductivity is driven by the anomalous Nernst effect. 
In fact, our calculations indicate a very large variation of the anomalous Nernst angle with energy, resulting also in a pronouced anomalous Nernst conductivity \cite{zimmermann14}. 

To estimate the magnitude of the spin current generated by the spin Nernst effect with the total SNC $\alpha_{\rm{tot}}$ of the order of 200 (A/K m) at room temperature, which is the case for Ag films with Pb and Cr impurities at the surface, in analogy to Ref.~\cite{tauber12}, we assume a sample size of 100$\times$100\,nm and an applied  temperature gradient of 50\,K/$\mu$m.  These parameters give us a value of the spin current of about 100\,$\mu$A. This is one order of magnitude larger than that obtained for bulk Cu(Au) alloy in Ref.~\cite{tauber12} by using the same computational approach.

Our observation that resonant scattering can lead to a giant SNE should have consequences in the presence of Kondo impurities at low temperature. 
It is well known that the Kondo resonance is very sharp \cite{hewson97}. 
Following the arguments presented here, Kondo impurities may strongly enhance the spin Nernst conductivity, just as they do with the thermopower \cite{boato67,andergassen11}.
However, its many-body-fluctuation character places the Kondo effect beyond the reach of density-functional methods that we use in the present work. 

To summarize, we demonstrated the possibility of drastic enhancement of the spin Nernst effect due to resonant impurity scattering taking place at surfaces of metallic films. As shown from a model and first principles calculations,
the presence of the resonant impurity states around the Fermi level results in a pronouncedly asymmetric behavior of longitudinal and/or spin Hall conductivites with energy, leading to large electrical and thermal contributions 
to the SNC.  
The reachable magnitude of the corresponding spin current that we predict opens new vistas in exploring the promises that the spin Nernst effect bares for spintronics applications. 
 
We would like to thank Martin Gradhand for fruithful discussions.
This work was financially supported by Deutsche Forschungsgemeinschaft projects MO 1731/3-1 and MA 4637/2-2 within the SPP 1538 Spin Caloric Transport.
We acknowledge computing time on the supercomputers JUQEEN and JUROPA at J\"ulich Supercomputing Center and JARA-HPC of RWTH Aachen University.

\end{document}